\documentclass[twocolumn,amsmath,amssymb]{snp}
\usepackage{graphicx}
\usepackage{dcolumn}
\usepackage{bm}
\def\be {\begin{equation}}
\def\ee {\end{equation}}
\def\bea {\begin{eqnarray}}
\def\eea {\end{eqnarray}}
\def\ra {\rightarrow}
\def\nn {\nonumber}

\newcommand{\vp}{\vec p}
\newcommand{\vq}{\vec q}

\newcommand{\mn}{\mu\nu}

\begin{document}
\title{In search of quark gluon plasma in nuclear collisions}
\author{Jan-e Alam}

\affiliation{Theoretical Physics Division, Variable Energy Cyclotron Centre, 
1/AF, Bidhan Nagar, Kolkata - 700064, India}
\maketitle
\begin{center}
{\bf Abstract}\\
\end{center}
At high temperatures and densities the nuclear matter undergoes  
a phase transition to a new state of matter called quark gluon plasma
(QGP). This new state of matter which existed in the  universe 
after a few microsecond of the big bang can be created 
in the laboratory by colliding two nuclei at relativistic energies. 
In this presentation we will discuss how the 
the properties of QGP can be extracted by analyzing 
the spectra of photons, dileptons and heavy flavours produced in nuclear
collisions at Relativistic Heavy Ion Collider (RHIC) and Large Hadron
Collider (LHC) energies. 

\section{Introduction} 
The theory of strong interaction - Quantum Chromodynamics (QCD) has a 
unique feature  - it possess the property of asymptotic freedom which
implies that at very high temperatures and/or  densities  nuclear matter
will  convert to a  deconfined state of quarks and gluons~\cite{collins}.
Recent lattice QCD based calculations~\cite{lqcd} indicate that the value
of the temperature for the nuclear matter to QGP transition $\sim 175$ MeV.
It is expected that such high temperature can be achieved 
in the laboratory by colliding nuclei at RHIC and LHC  energies. 

A high multiplicity system of deconfined  quarks and gluons with
power law type of momentum distributions
can created just after the nuclear collisions at high energies.
Interactions among these constituents may alter the 
momentum distribution of quarks and gluons from a power law to an 
exponential one - resulting in a thermalized state of quarks and gluons
with initial temperature, $T_i$. 
This thermalized system with high internal pressure  
expands very fast  as a consequence it cools and reverts to hadronic
matter at a temperature, $T_c\sim 175$ MeV.  
The hadrons formed after the hadronization of quarks
may maintain thermal equilibrium among themselves until the expanding
system becomes too dilute to support collectivity at a temperature,
$T_F (\sim 120$ MeV) called freeze-out temperature from where  
the hadrons fly freely from the 
interaction zone to the detector.

The electromagnetic (EM) probes~\cite{mclerran} (see ~\cite{rapp,alam1,alam2}
for review) {\it i.e.}  real photons and dileptons can be used to follow 
the evolution 
of the system from the pristine partonic  stage to the final hadronic
stage through an intermediary  phase transition or cross over. 
In the state of QGP some of the symmetries of the physical vacuum 
may either be restored or broken - albeit transiently. 
The electromagnetic  probes, especially the lepton pairs
can be used very effectively to investigate whether these symmetries
in the system  are restored/broken at any stage of the evolving matter. 
Results from theoretical calculations will be shown in the 
presentation to demonstrate this aspect of the electromagnetically 
interacting probes.  We will demonstrate that lepton pairs  can be
used very effectively to probe the collective motion (radial and elliptic)
of the system.

The other promising probe of the QGP that will be discussed here - 
is the depletion  of the transverse momentum 
spectra of energetic quarks (and gluons)  
in QGP. The magnitude of the depletion 
can be used to estimate the transport coefficients of 
QGP which is turn can be used to understand the fluidity 
of the matter. 

The  transport coefficients
of QGP and hot hadrons calculated by using perturbative QCD
and effective field theory respectively 
have been applied to evaluate the nuclear suppression ($R_{\mathrm AA}$)
of heavy flavours. Theoretical results on $R_{\mathrm AA}$ will
be compared with the experimental data available from RHIC
and LHC energies.  The azimuthal asymmetry of the system estimated 
through the single leptons originating from the decays of open heavy
flavours produced from the fragmentation of heavy quarks will also 
be discussed.

\section{The electromagnetic probes}
The dilepton production per unit four-volume 
from a thermal medium produced in heavy
ion collisions is well known to be given by:
\bea
\frac{dN}{d^4pd^4x}&=&-\frac{\alpha^2}{6\pi^3p^2}
L\left(M^2\right)f_{BE}(p_0) g^{\mu\nu}\,\nn\\
&&\times W_{\mu\nu}\left(p_0,\vp \right)
\label{eq1}
\eea
where the factor $L(M^2)=(1+{2m_l^2}/{M^2})~
(1-4m_l^2/M^2)^{1/2}~$ is of the order of unity for electrons,
$M(=\sqrt{p^2})$ being the invariant
mass of the pair and the hadronic tensor $W_{\mu\nu}$ is defined by
\be
W_{\mu\nu}(p_0,\vp)=\int d^4x\,e^{ip\cdot x}\langle[J^{em}_\mu(x),J^{em}_\nu(0)]\rangle
\label{eq2}
\ee
where $J^{em}_\mu(x)$ is the electromagnetic current and $\langle.\rangle$ indicates ensemble
average.
For a deconfined thermal medium such as the QGP, Eq.~(\ref{eq1})
leads to the
standard rate for lepton pair productions from
$q\bar q$ annihilation at lowest order.

The production of low mass dileptons 
from the decays of light vector mesons 
in the hadronic matter can be obtained as (see ~\cite{sabya}
for details): 
\bea
\frac{dN}{d^4pd^4x}&=&-\frac{\alpha^2}{6\pi^3p^2}
f_{BE}(p_0) g^{\mn}\nn\\
&&\times \sum_{R=\rho,\omega,\phi}K_R \rho^R_{\mn}(p_0,\vp)
\label{eq2}
\eea
where $f_{BE}$ is the thermal distributions for bosons,
 $\rho^R_{\mn}(q_0,\vq)$ is the spectral function of the 
vector meson $R (=\rho,\omega,\phi)$ in the medium, $K_R=F_R^2 m_R^2$,
$m_R$ is the mass of $R$ and $F_R$  is related to the decay
of $R$ to lepton pairs. 

The interaction of the vector mesons with the hadrons in the
thermal bath will shift the location of both the pole and the branch
cuts of the spectral function - resulting in mass modification or
broadening - which can be detected through the 
dilepton measurements and may be connected with 
the restoration of chiral symmetry in the thermal bath. In the present work the interaction of 
$\rho$ with thermal $\pi$,$\omega$, $a_1$, $h_1$~\cite{sabya,sabya2} 
and nucleons ~\cite{ellis}
have been considered to evaluate the in-medium spectral function  of
$\rho$. The finite temperature 
width of the $\omega$ spectral function has been taken from 
~\cite{Weise}.

To evaluate the dilepton yield
from a dynamically evolving system produced in heavy ion collisions (HIC)
one needs to 
integrate the fixed temperature production rate given by Eq.~\ref{eq2} 
over the space time evolution of the system  - from the initial
QGP phase to the final hadronic freeze-out state through a phase transition
in the intermediate stage.  
We assume that the matter is formed in QGP phase with 
zero net baryon density at temperature $T_i$ in HIC.
Ideal relativistic hydrodynamics with boost invariance~\cite{bjorken}
has been applied to study the evolution of the system. 

The EoS  required to close the hydrodynamic equations  is constructed 
by  taking results from lattice  QCD for high $T$ 
~\cite{lqcd} and hadron resonance gas comprising of
all the hadronic resonances up to mass of $2.5$ GeV ~\cite{victor,bmja}
for lower $T$. 
The system is assumed to get out of chemical equilibrium at
$T=T_{ch}=170$ MeV~\cite{tsuda}.  
The kinetic freeze-out temperature $T_{F}=120$ MeV 
fixed from the $p_T$ spectra of the produced hadrons.


\subsection{Invariant mass spectra of lepton pairs}
\begin{figure}
\centerline{\includegraphics[scale=0.3]{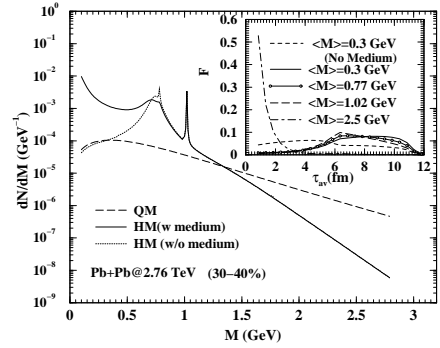}}
\caption{Invariant mass distribution of dileptons from 
hadronic matter (HM) for modified and unmodified  $\rho$ meson.
}
\label{fig1}
\end{figure}

The $M$ distribution of lepton pairs
originating from quark matter (QM) and hadronic matter (HM) with and
without medium effects on the spectral functions 
of $\rho$ and $\omega$ 
are displayed in Fig.~\ref{fig1}. 
We observe that for $M\,>\,M_{\phi}$
the QM contributions dominate. For  $M_{\rho}\lesssim M\lesssim M_{\phi}$
the  HM shines brighter than QM. For $M\,<M_{\rho}$, the HM (solid 
line) over shines
the QM due to the enhanced contributions primarily from the  medium
induced broadening of $\rho$ spectral function. However, the contributions
from QM and HM become comparable in this $M$ region if the medium
effects on $\rho$ spectral function is ignored (dotted line). 
Therefore, the results depicted in Fig.~\ref{fig1} indicate that a suitable choice of $M$ window will
enable us to unravel the contributions from a particular phase
(QM or HM). 
An appropriate choice of $M$ window 
will also allow us to extract the  medium induced effects. 

To further quantify these points we evaluate
the following~\cite{payalv2}:
\begin{eqnarray}
&&F=
\frac{\int^\prime  
\left(\frac{dN}{d^4xd^2p_TdM^2dy}\right)dxdyd\eta\tau d\tau d^2p_TdM^2}
{\int\left(\frac{dN}{d^4xd^2p_TdM^2dy}\right)dxdyd\eta\tau d\tau d^2p_TdM^2}\nn
 \label{eq3}
\end{eqnarray}
where the $M$ integration in both the numerator and 
denominator are performed  for selective 
windows from $M_1$ to $M_2$ with  mean $M$
defined as $\langle M\rangle = (M_1+M_2)/2$. 
While in the denominator 
the integration is done over the entire lifetime,
the prime in $\int^\prime$ in the numerator indicates 
that the $\tau$ integration
in the numerator is done from $\tau_1=\tau_i$ to $\tau_2=\tau_i+\Delta\tau$ 
with incremental $\Delta\tau$ until $\tau_2$ attains the life time of the
system.
In the inset of Fig.~\ref{fig1} 
$F$ is plotted against $\tau_{\mathrm av} (=(\tau_1+\tau_2)/2)$.
The results substantiate that pairs with high $\langle M\rangle\sim 2.5$ GeV 
originate from early time ($\tau_{\mathrm av}\lesssim 3$ fm/c, QGP phase) 
and pairs with 
$\langle M\rangle\sim 0.77$ GeV mostly emanate from late hadronic
phase ($\tau_{\mathrm av}\geq 4$ fm/c).  
The change in the properties of $\rho$ due to its interaction
with thermal hadrons in the bath 
is also visible through $F$ evaluated for $\langle M\rangle\sim 0.3$ GeV 
with and without medium effects. 

\begin{figure}
\centerline{\includegraphics[scale=0.35]{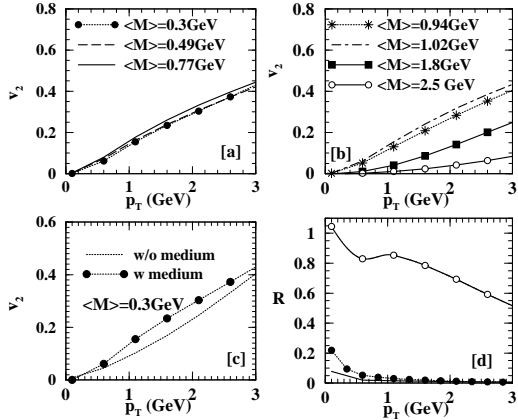}}
\caption{[a]  and [b] indicate elliptic flow of lepton pairs
as a function of $p_T$ for various $M$ windows.  [c] displays the effect
of the broadening of $\rho$ spectral function on the elliptic flow
for $\langle M\rangle = 300$ MeV.
[d] shows the variation of $R$  (see text)
with $p_T$ for $\langle M\rangle=0.3$ GeV, 0.77 GeV and 2.5 GeV.
}
\label{fig2}
\end{figure}

\begin{figure}
\centerline{\includegraphics[scale=0.35]{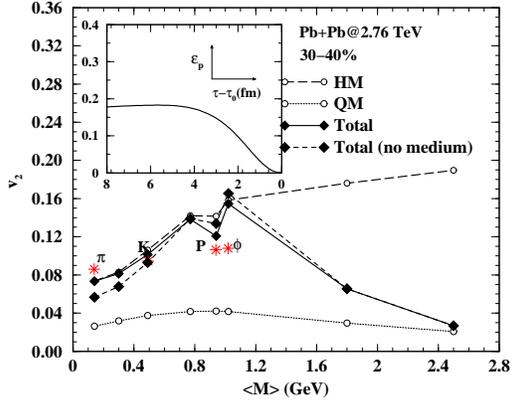}}
\caption{Variation of dilepton elliptic flow as
function of $\langle M\rangle$ for QM, HM (with and without medium effects) and
for the entire evolution. The inset shows the variation of momentum space anisotropy
with proper time.}
\label{fig3}
\end{figure}

\begin{figure}
\centerline{\includegraphics[scale=0.3]{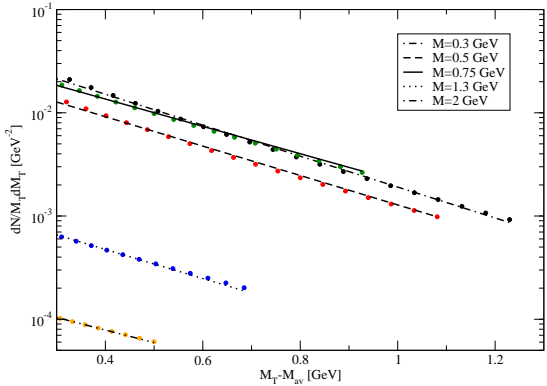}}
\caption{The dilepton yield plotted against $M_T-M_{av}$ for different $M$
windows for LHC initial condition.}
\label{fig4}
\end{figure}
\subsection{Elliptic flow of lepton pairs}

The elliptic flow of dilepton, $v_2(p_T,M)$ can be defined as:
\begin{eqnarray}
&&v_2=
\frac{\sum\int cos(2\phi)
\left(\frac{dN}{d^2p_TdM^2dy}\arrowvert_{y=0}\right) d\phi}
{\sum\int\left(\frac{dN}{d^2p_TdM^2dy}\arrowvert_{y=0}\right)d\phi }
 \label{eqv2}
\end{eqnarray}
where the $\sum$ stands for summation over
QM and HM phases.

Fig.~\ref{fig2} ([a] and [b]) show the differential elliptic
flow, $v_2(p_T)$ of dileptons arising from various $\langle M\rangle$
domains. We observe that for  $\langle M\rangle=2.5$ GeV $v_2$ is
small for the entire $p_T$ range
because these pairs arise from the early epoch (see inset of
Fig.~\ref{fig1}) when the flow is not developed entirely.
However, the $v_2$ is large for $\langle M\rangle=0.77$ GeV
as these pairs originate predominantly from the late hadronic
phase when the flow is fully developed.
It is also interesting to note that the medium induced enhancement
of $\rho$ spectral function provides a visible modification
in $v_2$ for dileptons below $\rho$ peak (Fig.~\ref{fig2} [c]).
The medium-induced effects lead to an enhancement of $v_2$
of lepton pairs which is culminating from the `extra' interaction
(absent when a vacuum $\rho$ is considered) of
the $\rho$ with other thermal hadrons in the bath.
In Fig.~\ref{fig2} [d] we depict the variation of  $R$ with $p_T$ for
$\langle M\rangle=0.3$ GeV (solid circle) 0.77 GeV (solid line)
and 2.5 GeV (open circle), the quantity $R$ is defined as
$R=v_2^{\mathrm QM}/(v_2^{\mathrm QM}+v_2^{\mathrm HM})$
where $v_2^{\mathrm i}$ is the elliptic flow of the phase $i(=QM+HM$. 
The results clearly illustrate
that $v_2$ of lepton pairs in the large $\langle M\rangle$ domain
originate from QM.

Fig.~\ref{fig3} shows $p_T$ integrated elliptic flow,
$v_2(\langle M\rangle)$ evaluated
for different $\langle M\rangle$ windows defined above.
The $v_2$ (which is proportional
to momentum space anisotropy, $\epsilon_p$) of QM is small because
the pressure gradient is not fully developed in the QGP phase
as evident from the inset plot of $\epsilon_p$ with $\tau$.
The hadronic phase $v_2$ has a peak around $\rho$ pole indicating
large flow at late times.
For $\langle M\rangle\>>\, m_\phi$ the $v_2$ obtained from the
combined phases approach the value corresponding to the $v_2$
for QGP.  Therefore, measurement of $v_2$ for large $\langle M\rangle$
will bring information of the properties of the QGP.
It is important to note that the $p_T$ integrated $v_2(\langle M\rangle)$
of lepton pairs with $\langle M\rangle\,\sim m_\pi, m_K$ is close to
the hadronic $v_2^\pi$ and $v_2^K$ if the thermal effects on $\rho$
properties are included. Exclusion of medium effects give lower
$v_2$ for lepton pairs compared to hadrons.
We also observe that the variation of $v_2(\langle M\rangle)$
with $\langle M\rangle$ has a structure similar to $dN/dM$ vs $M$.
As indicated by Eq.~\ref{eq1} we can write
$v_2(\langle M\rangle)\sim \sum v_2^{\mathrm i}\times f_{\mathrm i}$.
The structure of $dN/dM$ is reflected in $v_2(\langle M\rangle)$ through $f_i$.

\subsection{Radial flow of dileptons}
\begin{figure}
\centerline{\includegraphics[scale=0.3]{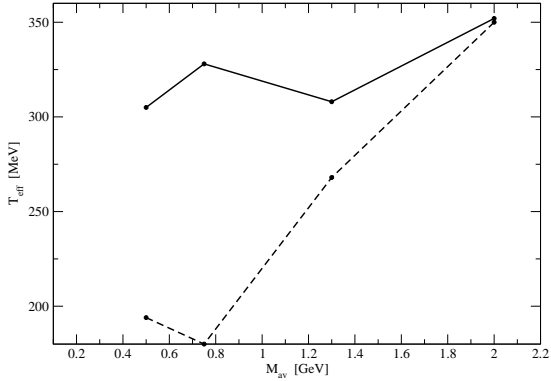}}
\caption{$T_{eff}$ for different values of the $M$-bins for LHC conditions.
The dashed line is obtained by setting $v_T=0$.}
\label{fig5}
\end{figure}

The transverse mass distributions of
the lepton pairs at LHC is displayed in Fig.~\ref{fig4}.
The variation of inverse slope (deduced from the from the
transverse mass  distributions of lepton pairs, Fig.~\ref{fig4}) 
with $\langle M\rangle$ for LHC is depicted in Fig.~\ref{fig5}.
The radial flow in the system is responsible for the 
rise and fall of  $T_{\mathrm eff}$ with $\langle M\rangle$ 
(solid line) in the mass region  ($0.5<$ M(GeV)$<1.3$),
for $v_T=0$ (dashed line) a completely different behaviour
is obtained. 

\begin{figure}
\centerline{\includegraphics[scale=0.35]{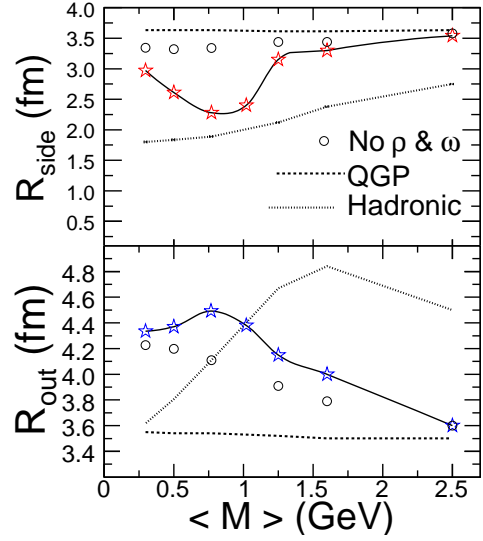}}
\caption{$R_{side}$ and $R_{out}$ as a function of
$\langle M \rangle$. The dashed, dotted and
the solid line (with asterisk) indicate the HBT radii
for the QGP, hadronic and
total dilepton contributions from all the phases respectively.
The solid circles are obtained by
switching off the contributions from $\rho$ and $\omega$.
}
\label{fig6}
\end{figure}
\subsection{Radial flow from HBT interferometry of lepton pairs}

It was shown in Ref.~\cite{payal} that the variation of HBT radii
($R_{side}$ and $R_{out}$) extracted from the correlation of dilepton pairs 
with $\langle M\rangle$ can used to extract collective properties of 
the evolving QGP. 
While the radius ($R_{\mathrm side}$) corresponding to $q_{side}$ 
is closely related to the transverse size of the system and 
considerably affected by the collectivity, 
the radius ($R_{\mathrm out}$) corresponding to $q_{out}$ measures both the 
transverse size and duration of particle emission. 
The extracted $R_{\mathrm side}$ and $R_{\mathrm out}$ for different 
$\langle M\rangle$ are shown in Fig.~\ref{fig6}.
The $R_{\mathrm side}$ shows non-monotonic dependence on $M$,
starting from a value close to QGP value (indicated by the dashed line)
it drops with increase in $M$ finally again approaching
the QGP value for $\langle M\rangle \,>\,m_\phi$. It can be shown
that $R_{side}\sim 1/(1+E_{\mathrm collective}/E_{\mathrm thermal})$.
In the absence of radial flow, $R_{\mathrm side}$ is independent of 
$q_{\mathrm side}$. With the radial expansion of the system
a rarefaction wave moves
toward the center of the cylindrical geometry as a consequence the radial
size of the emission zone decreases with time. 
Therefore, the size of the emission zone is larger at early times   
and smaller at late time. 
The high $\langle M\rangle$ regions 
are dominated by the early partonic phase 
where the collective flow has not been developed fully 
{\it i.e.} the ratio of collective to thermal energy is small
hence show larger $R_{\mathrm side}$ for the source.
In contrast, the lepton pairs with $M\sim m_\rho$ 
are emitted from the late hadronic phase where the size of the emission zone 
is smaller due to larger collective flow giving rise to
a smaller $R_{\mathrm side}$. The ratio of collective to thermal 
energy for such cases is quite large, which is reflected as a dip
in the variation of $R_{\mathrm side}$ with $\langle M\rangle$ 
around the $\rho$-mass region (Fig.~\ref{fig6} upper panel). 
Thus the variation of $R_{\mathrm side}$
with $M$ can be used as an efficient tool to measure the
collectivity in various phases of matter. 
The dip in $R_{\mathrm side}$ at $\langle M\rangle\sim
m_\rho$ is due to the contribution dominantly from the hadronic phase.
The dip, in fact vanishes if the contributions from $\rho$ and $\omega$ 
is switched off (circle in Fig.~\ref{fig6}). 
We observe that by keeping the $\rho$ and $\omega$ contributions
and setting radial velocity, $v_r=0$, the dip in $R_{\mathrm side}$
vanishes, confirming
the fact that the dip is caused by the radial flow of the hadronic matter.   
Therefore, the value of 
$R_{\mathrm side}$ at $\langle M\rangle\sim m_\rho$ may be used
to estimate the average $v_r$ in the hadronic phase.

The $R_{\mathrm out}$ probes both the transverse dimension and the 
duration of emission as a consequence unlike $R_{\mathrm side}$ it does not
remain constant even in the absence of radial flow and
its variation with $M$ is complicated. The large $M$ regions are
populated by lepton pairs from early partonic phase where the
effect of flow is small and the duration of emission is also
small - resulting in smaller values of $R_{\mathrm out}$. 
For lepton pair from $M\sim m_\rho$ the flow is large
which could have resulted in a dip as in $R_{\mathrm side}$ in
this $M$ region. However, $R_{\mathrm out}$ probes the duration 
of emission too which is large for hadronic phase.
The larger duration compensates the reduction of $R_{\mathrm out}$ 
due to flow in the hadronic phase 
resulting in a bump in $R_{\mathrm out}$ for $M\sim m_\rho$
(Fig.~\ref{fig6} lower panel). Both $R_{\mathrm side}$ and $R_{\mathrm out}$
approach QGP values for $\langle M\rangle\sim 2.5$ GeV implying dominant
contributions from partonic phase. 

\begin{figure}
\centerline{\includegraphics[scale=0.35]{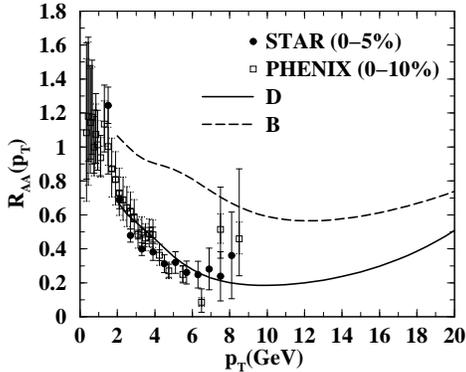}}
\caption{$R_{AA}$ as a function of  $p_T$ for $D$ and $B$ mesons  at LHC.
Experimental data taken from ~\cite{ALICE}.}
\label{fig7}
\end{figure}

\section{Suppression of heavy flavours in QGP}
The depletion of  hadrons with high transverse momentum ($p_T$)
produced in Nucleus + Nucleus collisions with respect to those
produced in proton + proton (pp) collisions has been considered
as a signature of QGP formation.
The two main processes which cause the depletion are
(i) the elastic collisions
and (ii) the radiative loss or the inelastic collisions 
of the  high energy partons with the quarks,
anti-quarks and gluons in the thermal bath.

In the present work we focus on the energy loss of 
heavy quarks in QGP in deducing the properties of the medium. 
Because (i) the abundance of charm and bottom quarks in the partonic plasma
for the expected range of temperature to be attained in the experiments 
is small, consequently the bulk properties of the plasma is not decided 
by them and (ii) they produce early and therefore, can witness the
entire evolution history.
Hence heavy quarks may act as an efficient probe for the 
diagnosis of QGP. 
The depletion of  heavy quarks in QGP 
has gained importance recently in view of the measured
nuclear suppression in the $p_T$ spectra of 
non-photonic single electrons~\cite{stare,phenixe}.

We assume here that  the light quarks and gluons thermalize 
before heavy quarks. The charm and bottom quarks
execute Brownian motion~\cite{we1} (see references therein)
in the heat bath of QGP.
Therefore, the interaction of the heavy quarks with 
QGP may be treated as the 
interactions between equilibrium and non-equilibrium degrees
of freedom. 
The Fokker-Planck (FP) equation provide an appropriate framework for
the evolution of the heavy quark in the expanding
QGP heat bath which can be written as~\cite{we1}:
\bea
\frac{\partial f}{\partial t}&=&C_{1}(p_{x},p_{y},t)\frac{\partial^{2}f}{\partial p_{x}^{2}}~+C_{2}(p_{x},p_{y},t)
\frac{\partial^{2}f}{\partial p_{y}^{2}}\nn\\&+&~C_{3}(p_{x},p_{y},t)\frac{\partial f}{\partial p_{x}}~+C_{4}(p_{x},p_{y},t)
\frac{\partial f}{\partial p_{y}}\nn\\&+&~C_{5}(p_{x},p_{y},t)f~+C_{6}(p_{x},p_{y},t).
\label{fpeqcartesian}~~.
\eea
where,
\bea
C_{1}&=& D\\C_{2}&=& D\\
C_{3}&=& \gamma ~p_{x}~+2~\frac{\partial D}{\partial p_{T}}~
\frac{p_{x}}{p_{T}}\\C_{4}&=& \gamma ~p_{y}~+2~\frac{\partial D}{\partial p_{T}}~\frac{p_{y}}{p_{T}}
\\C_{5}&=& 2~\gamma ~+\frac{\partial \gamma }{\partial p_{T}}~\frac{p_{x}^{2}}{p_{T}}~+
\frac{\partial \gamma }{\partial p_{T}}~\frac{p_{y}^{2}}{p_{T}}\\C_{6}&=& 0~~.
\eea
where the momentum, $\textbf{p}=(\textbf{p}_T,p_z)=(p_x,p_y,p_z)$, $\gamma$ is the
drag coefficient and $D$ is the diffusion coefficient. 
We numerically solve Eq.~\ref{fpeqcartesian}~\cite{antia} with the boundary conditions:
$f(p_x,p_y,t)\ra 0$ for $p_x$,$p_y\ra \infty$ and  the initial (at time $t=\tau_i$)
momentum distribution of charm and bottom quarks are taken MNR code~\cite{MNR}.

The system under study has two components. 
The equilibrium component, the  QGP comprising of the
light quarks and the gluons.
The non-equilibrium
component, the heavy quarks produced due to the collision
of partons of the colliding nuclei has momentum distribution
determined by the perturbative QCD (pQCD), 
which  evolves due to their 
interaction with the expanding QGP background. 
The evolution of the heavy quark momentum distribution 
is governed by the FP equation.
The interaction of the heavy quarks with the QGP
is contained in the drag and diffusion coefficients.
The drag and diffusion coefficients are provided as inputs, which are,
in general, dependent on both temperature and momentum. 
The evolution of the temperature of the background 
QGP with time is governed by 
relativistic hydrodynamics. The solution of the FP equation
at the (phase) transition point for the charm and bottom quarks
gives the (quenched) momentum distribution of hadrons ($B$ and $D$ mesons) 
through fragmentation. The fragmentation of the initial momentum
distribution of the heavy quarks results in the unquenched 
momentum distribution of the $B$ and $D$ mesons.  The ratio
of the quenched to the unquenched $p_T$ distribution is the
nuclear suppression factor which is experimentally measured.
The quenching is due to the dragging of the heavy quark by QGP.
Hence the properties  of the QGP can be extracted
from the suppression factor. 

\subsection{Nuclear suppression factor}

The variation of the nuclear suppression factor, $R_{AA}$
~\cite{we1} with $p_T$ of the electron originating from the decays of
$D$ and $B$ mesons have been displayed in Fig.~\ref{fig7}
for RHIC initial condition ($T_i=300$ MeV). 
A less suppression of $B$ is observed compared to $D$.
The theoretical results show a slight upward 
trend for $p_T$ above 10 GeV both for mesons containing 
charm and bottom quarks.  Similar trend has recently been experimentally
observed for light mesons at LHC energy~\cite{CMS}.
This may originate from the fact that 
the drag (and hence the quenching) for charm and 
bottom quarks are less at higher momentum. 

\begin{figure}
\centerline{\includegraphics[scale=0.35]{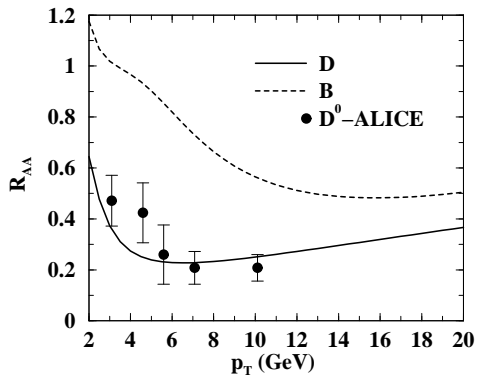}}
\caption{$R_{AA}$ as a function of  $p_T$ for $D$ and $B$ mesons  at LHC.
Experimental data taken from ~\cite{ALICE}.}
\label{fig8}
\end{figure}

The same formalism is extended to evaluate the nuclear suppression 
factor, $R_{AA}$ both for charm and bottom at LHC energy. Result has been 
compared with the recent ALICE data(Ref.~\cite{ALICE}) in Fig.~\ref{fig8}.
The data is reproduced well by assuming formation of QGP 
at an initial temperature $\sim 550$ MeV after Pb+Pb collisions at 
$\sqrt{s_{\mathrm NN}}=2.76$ TeV.  

\subsection{Elliptic flow of heavy flavours}
\begin{figure}
\centerline{\includegraphics[scale=0.35]{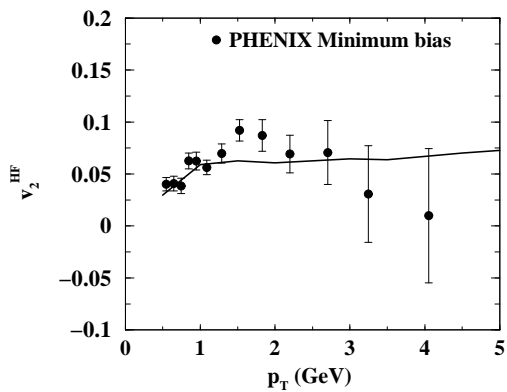}}
\caption{Elliptic flow of single electrons originating from the
heavy mesons decays.
}
\label{fig9}
\end{figure}

In Fig.~\ref{fig9} we compare the experimental data obtained by  the
PHENIX~\cite{phenixemb} collaborations 
for Au + Au minimum bias collisions at $\sqrt{s_{\mathrm NN}}=200$ GeV with
theoretical results obtain in the present work. We observe that the 
data can be reproduced by including both radiative and collisional loss
with $c_s=1/\sqrt{4}$.
In this case $v_2^{HF}$ first increases  and reaches a maximum of about 7\% 
then  saturates for  $p_T>2$ GeV. 
However, with ideal equation of state ($c_s^2=1/3$)  we fail to reproduce data. 
This is because with larger value of $c_s$ the system expands faster 
as a result has shorter life time for fixed $T_i$ and $T_c$. 
Consequently the heavy quarks get lesser time to interact with the expanding
thermal system and  fails to generate enough flow.
From the energy dissipation
we have evaluated the shear viscosity ($\eta$) to entropy ($s$) density ratio
using the relation~\cite{mmw}: $\eta/s\sim 1.25T^3/\hat{q}$, where $\hat{q}=
\langle p^2_T\rangle/L$ and $dE/dx\sim \alpha_s\langle p^2_T\rangle$
~\cite{RB}, $L$ is the length traversed by the heavy quark.
The average value of $\eta/s\sim 0.1-0.2$,
close to  the AdS/CFT bound~\cite{KSS}.

\section{Summary}
 In this work we have discussed the productions
of lepton pairs from nuclear collisions at relativistic energies
and shown that lepton pairs can trace the evolution of collectivity 
of the system. 
The elliptic flow and the nuclear suppression factor 
of the electrons originating
from the heavy flavour decays have been 
studied by 
including both the radiative and the collisional processes of energy loss
in evaluating the effective drag and diffusion coefficients of the heavy quarks.
The results have been compared with the available experimental data
and properties of QGP expected to be formed at RHIC collisions 
have been extracted.

{\bf Acknowledgment:} The author is grateful to 
Trambak Bhattacharyya, Santosh K Das, Sabyasachi Ghosh, Surasree
Mazumder, Bedangadas Mohanty,
Payal Mohanty and Sourav Sarkar for collaboration and
to Tetsufumi Hirano for providing hadronic chemical potentials.

\end{document}